\documentclass[twocolumn,prl,showpacs]{revtex4}
\usepackage{epsfig}
\usepackage{graphics}
\usepackage{graphicx}
\usepackage{dcolumn}
\usepackage{bm}

\begin{document}

\title{Evolution of Coherence and Superconductivity in Electron-Doped Cuprates}
\author{M. M. Qazilbash$^{1,2}$, B. S. Dennis$^1$, C. A. Kendziora$^3$, Hamza Balci$^2$, R. L. Greene$^2$, and G. Blumberg$^{1,\dag}$}
\affiliation{
$^{1}$Bell Laboratories, Lucent Technologies, Murray Hill, NJ 07974 \\
$^{2}$Center for Superconductivity Research, Department of Physics, University of Maryland, College Park, MD 20740\\
$^{3}$United States Naval Research Laboratory, Code 6365, Washington D.C. 20375}

\date{\today}

\begin{abstract}
The electron-doped cuprates were studied by electronic Raman spectroscopy across the entire region of the superconducting (SC) phase diagram.
We determined that the magnitude of the SC order parameter varies between 4.6 and 3.5 $k_BT_c$, consistent with weak coupling BCS theory.
Using a ``Raman conductivity" sum rule, we found that doped carriers divide into coherent quasi-particles (QPs) and carriers that remain incoherent.
The coherent QPs mainly reside in the vicinity of ($\pm \pi/2a$, $\pm \pi/2a$) regions of the Brillouin zone.
The carriers doped beyond optimal doping remain incoherent.
Only coherent QPs contribute to the superfluid density in the SC state.
\end{abstract}


\maketitle

{\em Introduction}.--
In recent years, there has been renewed interest in the physical properties of electron \emph{n}-doped
high T$_c$ superconducting (SC) cuprates. While there is
considerable evidence that the SC order parameter (OP) of the hole
\emph{p}-doped cuprates has $d$-wave symmetry in the entire doping
range of their SC phase \cite{Harlingen,kirtleyreview}, the
situation for the \emph{n}-doped compounds still remains
controversial.
Earlier Raman data for optimally-doped (OPT) samples implied a non-monotonic $d$-wave OP
\cite{BlumbergNCCO}.
There is disagreement among the few experiments that have studied the doping dependence of the SC OP \cite{Amlan,skinta1,skinta2,prozorov2}.
Angle-resolved photoemission spectroscopy (ARPES) data at optimal doping indicates the presence of both well-defined quasi-particles (QPs), as well as ill defined incoherent excitations \cite{Armitage:2002}.
There is current interest in examining the relationship between the coherence properties of introduced carriers and development of the SC OP with doping.
We report a systematic low energy electronic Raman spectroscopy
(ERS) study of Pr$_{2-x}$Ce$_x$CuO$_{4-\delta}$ (PCCO) and
Nd$_{2-x}$Ce$_x$CuO$_{4-\delta}$ (NCCO) single crystals and films
with different cerium doping covering the entire SC region of the
phase diagram and determine the magnitude of the OP as a function
of doping.
By applying a "Raman conductivity" sum rule
\cite{Sriram,BlumbergRamanDrude}, we find that carriers doped
beyond optimal doping remain incoherent and do not contribute to
the superfluid density.

{\em Experimental}.-- Raman scattering was performed from natural
$ab$ surfaces of single crystals and films of PCCO and single
crystals of NCCO. Crystals with different Ce concentrations were
grown using a flux method as described in ~\cite{JPeng}. After
growth, the crystals were annealed in an Ar-rich atmosphere to
induce superconductivity. The SC transitions were measured by a
SQUID magnetometer. The Ce concentration of the crystals was
measured with wavelength dispersion spectroscopy. \emph{c}-axis
oriented PCCO films were grown on strontium titanate substrates
using pulsed laser deposition \cite{Maiser}. The films were grown
to a thickness of about 0.8 to 1 $\mu$m to minimize the substrate
contribution to the Raman signal. The SC transitions were measured
by \emph{ac} susceptibility. The films provide an opportunity to
study the extremes of the SC phase because of better control of Ce
doping in under-doped (UND) and highly over-doped (OVD) samples.

The samples were mounted in an optical continuous helium flow
cryostat. The spectra were taken in a pseudo-backscattering
geometry using linearly polarized excitations at 647 and 799 nm
from a Kr$^+$ laser. Incident laser powers between 0.5 and 4~mW
were focused to a $50 \times 100$~$\mu$m spot on the sample
surface. The spectra were measured at temperatures between 4 and
30~K by a custom triple grating spectrometer and the data were
corrected for instrumental spectral response.
The sample temperatures quoted in this work have been corrected for laser heating.

{\em Raman scattering symmetries}.-- The polarization directions
of the incident, \textbf{e}$_i$, and scattered, \textbf{e}$_s$,
photons are indicated by (\textbf{e}$_i$\textbf{e}$_s$) with
x=[100], y=[010], x'=[110] and y'=[$\overline{1}$10]. The
presented data were taken in (xy), (x'y') and (xx) scattering
geometries. For the tetragonal $D_{4h}$ symmetry of the
\emph{n}-doped cuprates, these geometries correspond to
$B_{2g}$+$A_{2g}$, $B_{1g}$+$A_{2g}$ and $A_{1g}$+$B_{1g}$
representations. Using circularly polarized light we confirmed
that the contribution to the $A_{2g}$ channel is very weak for
both PCCO and NCCO.

The electronic Raman response function, $\chi''^{(is)}(\omega)$,
for a given polarization geometry (\textbf{e}$_i$\textbf{e}$_s$)
is proportional to the sum over the density of states at the Fermi
surface (FS) weighted by the square of the momentum \textbf{k}
dependent Raman vertex $\gamma_{\textbf{k}}^{(is)}$
\cite{Dierker&Hackl,Klein,Devereaux1}. Because the scattering
geometries selectively discriminate between different regions of
the FS, ERS provides information about both the magnitude and the
\textbf{k} dependence of the SC OP. In the effective mass
approximation $\gamma_{\bf{k}}^{B_{1g}} \propto t(\cos k_{x}a -
\cos k_{y}a)$ and $\gamma_{\bf{k}}^{B_{2g}} \propto 4t' (\sin
k_{x}a \sin k_{y}a)$ where $t$ and $t'$ are nearest and
next-nearest neighbor hopping integrals in the tight-binding
model. For the $B_{2g}$ channel the Raman vertex is maximum around
($\pi/2a$, $\pi/2a$) and equivalent regions of the Brillouin zone
(BZ) and vanishes along (0, 0)$\rightarrow$($\pi/a$, 0) and
equivalent lines. For the $B_{1g}$ channel nodal (0,
0)$\rightarrow$($\pi/a$, $\pi/a$) diagonals do not contribute to
the intensity that mainly integrates from the regions near
intersections of the FS and the BZ boundary.

\begin{figure}[t]
\epsfig{figure=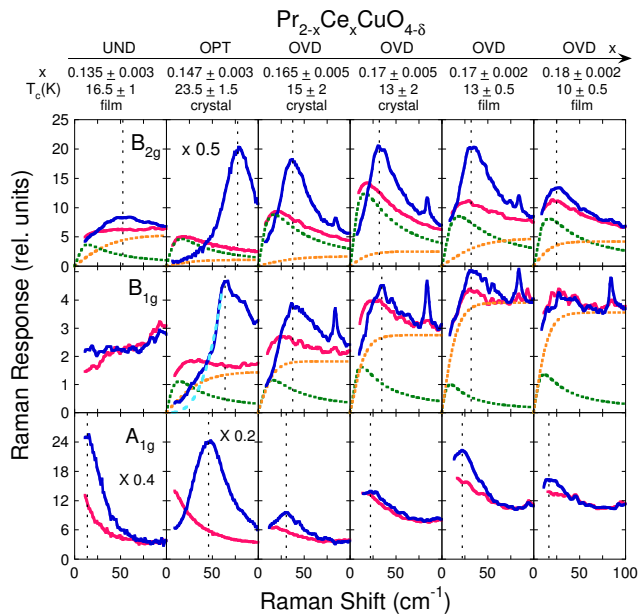,width=85mm}
\caption{Doping dependence of the low energy electronic Raman
response of PCCO single crystals and films for $B_{2g}$, $B_{1g}$
and $A_{1g}$ channels obtained with 647~nm excitation. The columns
are arranged from left to right in order of increasing cerium
doping. The light (red) curves are the data taken just above the
respective T$_c$ of the samples. The normal state response in the
$B_{2g}$ and $B_{1g}$ channels is decomposed into a coherent Drude
contribution in a quasi-elastic form (green dotted line) and an
incoherent continuum (yellow dotted line). The dark (blue) curves
show the data taken in the SC state at T~$\approx$~4~K. The dashed
vertical lines indicate positions of the 2$\Delta$ peak. For the
OPT crystal a low-frequency $\omega^3$ power law is shown in the
$B_{1g}$ panel for comparison (light-blue dotted line). OVD
crystals and films show similar behavior, indicating the good
quality of the films. }\label{Fig.1}
\end{figure}

\begin{figure}[t]
\epsfig{figure=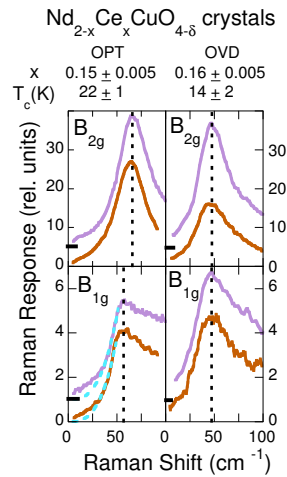,width=38mm}
\caption{A comparison of the Raman response in the SC state for
OPT (first column) and OVD (second column) NCCO crystals. The
first and second rows show spectra for $B_{2g}$ and $B_{1g}$
channels respectively. The violet (light) and orange (dark) curves
are data taken with red laser excitation ($\lambda_L$ = 647 nm)
and near-IR excitation ($\lambda_L$ = 799 nm). The $B_{2g}$ and
B$_{1g}$ data for $\lambda_L$ = 647 nm has been shifted up by 5
units and 1 unit respectively. All spectra are taken at T
$\approx$ 4~K. The dashed vertical lines show the positions of the
2$\Delta$ peaks. For the OPT crystal a low-frequency $\omega^3$
power law is shown in the $B_{1g}$ panel for comparison
(light-blue dotted line). } \label{Fig.2}
\end{figure}

Figure 1 exhibits the evolution with doping of the Raman response function above $T_c$ and at 4~K, deep in the SC state, for all three observable symmetry channels: $B_{2g}$, $B_{1g}$ and $A_{1g}$.
The total intensity is significantly stronger in the $B_{2g}$ than in the $B_{1g}$ channel for all doping concentrations.
This underlines the importance of the next-nearest neighbor
hopping $t'$ for the \emph{n}-doped cuprates and is in contrast
with \emph{p}-doped cuprates where the response in the $B_{2g}$
channel is generally weaker than in $B_{1g}$ \cite{Liu}. Although
Coulomb screening should lead to a much weaker Raman response in
the fully symmetric channel, we find that the intensities in the
$A_{1g}$ channel are of  the same order of magnitude as those in
crossed polarizations in the OVD samples and are significantly
stronger in the UND and OPT samples. This lack of screening is not
understood within the framework of existing theoretical models.

{\em Pair breaking excitations out of the SC condensate}.--
In the $B_{2g}$ and $A_{1g}$ channels, the pair-breaking 2$\Delta$ coherence peaks appear for all doping concentrations while  in the $B_{1g}$ channel the 2$\Delta$ peaks are negligibly weak in the UND and the most OVD films.
For the OPT crystal (T$_c$ = 23.5 K) the coherence peak energy is
larger for the $B_{2g}$ channel compared with that in $B_{1g}$ and
for all channels it is larger than the scattering rate $\Gamma$
obtained from the spectra in the normal state discussed in the
following section. The intensity below the coherence peaks
vanishes smoothly without a threshold to the lowest frequency
measured. The absence of a threshold that has been observed in
$s$-wave superconductors precludes interpretation in terms of a
fully gapped Fermi surface \cite{Dierker&Hackl}. The smooth
decrease in the Raman response below the 2$\Delta$ peak is
consistent with nodes in the gap. We compare the low-frequency
tail in the $B_{1g}$ response (Fig. 1) to an $\omega^3$ power law
that is expected for a $d_{x^{2}-y^{2}}$-wave superconductor in
the clean limit \cite{Devereaux2}. The observed deviation from
cubic to a linear response at the lowest frequencies is an
indication of low-energy QP scattering \cite{optics, Devereaux3}.
The data for OPT PCCO is very similar to that for OPT NCCO (see
Fig.~2) which was interpreted in terms of a non-monotonic $d$-wave
order parameter with nodes along the (0, 0)$\rightarrow$($\pi/a$,
$\pi/a$) diagonal and the maximum gap being closer to this diagonal than to the BZ boundaries \cite{BlumbergNCCO}.

Interestingly, in the OVD PCCO samples (Fig.~1) and OVD NCCO
crystal (Fig.~2), the 2$\Delta$ peak positions are at the same
energies for both the $B_{2g}$ and $B_{1g}$ channels.
The 2$\Delta$ peak positions and intensities decrease in the OVD regime compared
to the OPT samples. Moreover, $2\Delta \sim \Gamma$ indicates that
superconductivity is approaching the dirty or disordered
limit~\cite{Devereaux3}.
The Raman response below the 2$\Delta$-peaks vanishes smoothly and no
well-defined threshold is observed. The data for the 799~nm
excitation is measured down to 4.5~cm$^{-1}$ and shows no obvious
sub-gap threshold. The peak positions and the sub-gap Raman
response in the NCCO crystals are almost independent of the laser
excitation energies~(Fig.~2). A similar symmetry independent pair
breaking peak energy with continuously decreasing Raman scattering
intensity down to the lowest frequencies measured has been
observed in the Raman spectra in OVD samples of \emph{p}-doped
Bi-2212 \cite{holeoverdoped}. The Raman data presented in
Figs.~1-2 for OVD \emph{n}-doped samples is similar to the Raman
data for OVD Bi-2212. The coincidence of the coherence peak
energies in the $B_{1g}$ and $B_{2g}$ channels is probably caused
by enhanced QP scattering~\cite{Devereaux3}.
Nevertheless, the continuously decreasing Raman intensity below
the coherence peak is not inconsistent with a nodal gap structure.

Figure 3(a) summarizes the energy of the  $2\Delta$ coherence peak
as a function of doping for all three scattering channels. The
coherence peak energy has a pronounced maximum at optimal doping.
The 2$\Delta$ features in the $A_{1g}$ channel occur at lower
energies compared to those in the $B_{2g}$ and $B_{1g}$ channels.
For comparison, we include the value of twice the SC gap energy
obtained from point contact tunneling spectroscopy
\cite{Qazilbash}.
While for OPT and OVD samples the maximum value
of the Raman 2$\Delta$ peak positions are very similar to the single
particle spectroscopy gap values, this is not the case for the UND
samples where the tunneling spectroscopy data exhibits a gap that
is larger than the Raman coherence peaks.
For UND samples the two QPs excited out of the SC condensate by Raman processes continue to interact, binding into a collective excitonic
state that costs less energy than excitation of two independent
QPs. Similar observations were made previously for \emph{p}-doped
cuprates \cite{Liu,Blumberg:97}. The importance of the final state
interactions in the formation of a collective mode in UND cuprates
has been demonstrated in Refs. \cite{BardasisSchrieffer:61,Chubukov:99}.

Reduced energies of the coherence peaks, $2\Delta/k_{B}T_{c}$, are
plotted in Fig. 3b as a function of doping.
For the channel that exhibits the highest ratio, $B_{2g}$, the values fall between 4.6 for the UND and OPT samples and 3.5 for the most OVD samples, within
the prediction of the mean-field BSC values for \emph{d}-wave
superconductors \cite{Maki}.
The coherence peak energy remains below $4.2k_{B}T_{c}$ for the $B_{1g}$ channel and is even lower for the $A_{1g}$ channel, particularly for the UND sample.
The reduced energies for all the channels are significantly lower than for \emph{p}-doped materials \cite{Liu,Blumberg:97,Kendziora:95} suggesting a BCS weak coupling limit for the \emph{n}-doped cuprates.

\begin{figure}[t]
\epsfig{figure=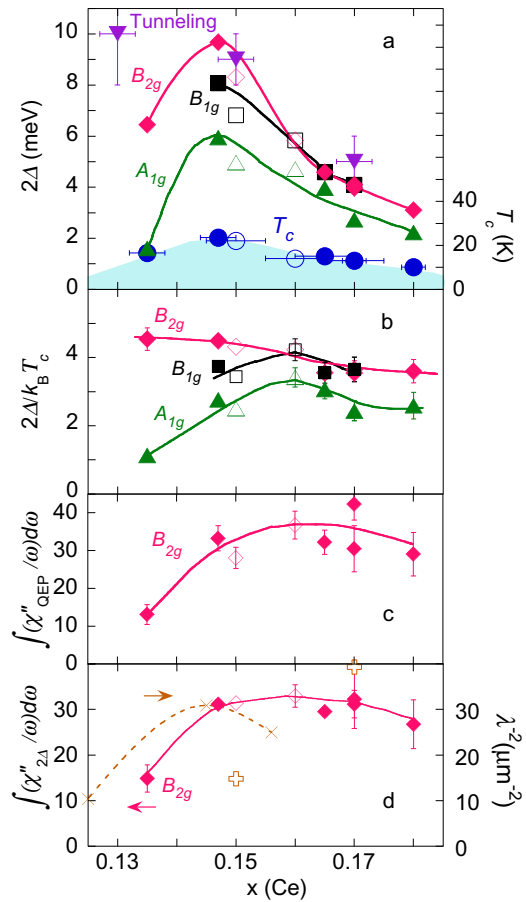,width=70mm}
\caption{The phase diagram of PCCO (filled symbols) and NCCO (open
symbols) superconductors explored by ERS. Panels show: (a) T$_c$
and 2$\Delta$ peak positions (gap magnitudes) for $B_{1g}$,
$B_{2g}$ and $A_{1g}$ channels as well as the distance between
coherence peaks from point contact tunneling spectroscopy
\cite{Qazilbash}; (b) The magnitude of the reduced gap
($2\Delta/k_{B}T_{c}$) for three Raman channels; (c) The
integrated QP (Drude) response just above $T_{c}$; (d) The
integrated intensity in the SC coherence peak at 4~K. For
comparison, we plot 1/$\lambda(0)^2$ values from refs.
~\cite{skinta1} ($\times$) and \cite{prozorov2} (open crosses).
Error bars on the cerium concentrations are shown only on the
T$_c$ data points to preserve clarity of the figure. Solid lines
are guides to the eye. } \label{Fig.3}
\end{figure}

{\em Evolution with carrier doping in the normal state}.-- In the
normal state the Raman response can be decomposed into a
featureless continuum and a defined low-frequency
quasi-elastic scattering peak (QEP): $\chi''_N(\omega) =
\chi''_{QEP}(\omega) + \chi''_{MFL}(\omega)$. The QEP response
$\chi''_{QEP}(\omega) = a^{(is)} \frac{\Gamma \omega}{\omega^2 +
\Gamma^2}$ is described in a Drude model as a well defined  QP contribution from doped carriers \cite{BlumbergRamanDrude,Andreas} while the featureless continuum
$\chi''_{MFL}(\omega) = b^{(is)} \tanh({\omega}/{\omega_c})$ represents a collective incoherent response \cite{Varma}.
Symmetry dependent $a^{(is)}$ and $b^{(is)}$ parameters control the spectral weight in these coherent and incoherent channels, $\omega_c$ is a cut-off frequency of order $k_{B}T$ \cite{Varma} and the QEP
position $\Gamma$ is the Drude scattering rate that at low temperatures remains between 2 and 2.5~meV for the entire studied doping range.
This deconvolution of the Raman response into two components presented here is consistent with the ARPES data that simultaneously displays defined QPs in the vicinity of the ($\pi/2a$, $\pi/2a$) point and ill defined excitations in the other parts of the FS \cite{Armitage:2002}.
One can observe from the deconvolution that the Raman response in the $B_{2g}$ channel is dominated by the QP (Drude) response while in the $B_{1g}$ channel by the incoherent continuum \cite{Andreas,films}.
This confirms that the defined QP states reside in the vicinity of the ($\pm \pi/2a$, $\pm \pi/2a$) regions of the BZ.

Figure 3(c) displays the evolution of the integrated QP spectral weight of the ``Raman conductivity" \cite{BlumbergRamanDrude}
$I_{N}^{B_{2g}}(x)=\int (\chi''^{B_{2g}}_{QEP} / {\omega}) \,d\omega$
as a function of electron (Ce) doping $x$.
While on the UND side $I_{N}^{B_{2g}}(x)$ exhibits the expected increase proportional to
$x$, the integrated coherent contribution saturates above optimal
doping $x \gtrsim 0.145$.
At these higher concentrations, additional carriers contribute only to the incoherent response and
can be observed as an increasing intensity of the featureless Raman continuum $\chi''_{MFL}(\omega)$,  particularly in the $B_{1g}$ channel (Fig.~1).

{\em SC coherence peak intensity and sum rules}.--
In Fig.~3(d) we plot the integrated coherence intensity in the SC state,
$I_{SC}^{B_{2g}}(x)=\int({\chi''^{B_{2g}}_{2\Delta}}/{\omega})\,d\omega$, where
$\chi''^{B_{2g}}_{2\Delta}(\omega)$ is the SC coherence response
with the background subtracted. We note that the values of the
integrated coherence intensity do not change from the normal to SC
state (Fig.~3c and d) demonstrating a sum rule for the ``Raman
conductivity". For the non-symmetric channels $I_{SC}^{(is)}
\propto \sum_{\bf{k}}{(\gamma_{\bf{k}}^{(is)})^2
\frac{{\Delta_{\bf{k}}}^2}{{2E_{\bf{k}}}^3}}$ is proportional to
the superfluid density $n_s \propto
\sum_{\bf{k}}{\frac{{\Delta_{\bf{k}}}^2}{{2E_{\bf{k}}}^3}}$
weighted by the square of the Raman coupling vertex.
Here $E_{\bf{k}}$ is the QP dispersion in the SC state.
The superfluid densities ($n_s\propto1/\lambda^2$) obtained from penetration depth ($\lambda$) measurements \cite{skinta1,prozorov2} are plotted in Fig.~3(d) for comparison.
The $I_{N}^{B_{2g}}(x)= I_{SC}^{B_{2g}}(x)$ equality implies that
not all doped electrons but only Drude QPs control the
superfluid density and that the incoherent carriers doped above
optimal doping do not contribute to the superfluid stiffness \cite{resonance}.

{\em Summary}.-- ERS has been investigated across the entire SC
phase diagram of the  \emph{n}--doped cuprates. The SC gap
magnitude has a maximum near optimal doping. The gaps measured by
ERS are in agreement with those measured by single particle
spectroscopy for OPT and OVD samples. For the UND film, the Raman
data shows a smaller gap implying strong final state interactions.
The reduced coherence peak values (2$\Delta/k_BT_c$) for the $B_{2g}$ channel decrease monotonically from 4.6 for the UND sample to 3.5 for OVD samples.
Using the ``Raman conductivity" sum rule, we find that carriers doped beyond optimal doping
remain incoherent and do not contribute to the Drude conductivity and superfluid density.
The reduced ratio of coherent on the background of incoherent carriers possibly explains the fragility
of superconductivity in the OVD \emph{n}-doped cuprates.

The authors thank A. Koitzsch, A. Gozar, Y. Dagan, C. P. Hill and
M. Barr for assistance, B. Liang for WDS measurements, V. N.
Kulkarni for Rutherford Backscattering on PCCO films, and Z. Y. Li
for crystal growth. This work was supported in part by NSF Grant
No. DMR 01-02350.


\begin{references}
\bibitem{optics} Optical conductivity data also indicates that OPT PCCO films are between
the clean and dirty limits \cite{Bontemps}.
\bibitem{films}
The featureless continuum is stronger for the film samples
compared to the single crystals. This is an indication of
additional disorder and/or possible strain fields.
\bibitem{resonance}
The above analysis assumes that the resonant properties of the Raman vertex, i.e. the optical
conductivity near the laser excitation of 647 nm (1.9 eV), do not change with doping in the relevant doping range (0.13 $<$ x $<$ 0.18). Indeed, there is only a weak change in optical conductivity
above 0.8 eV with doping in the relevant doping range~\cite{arima}.
\end{references}
\end{document}